# Workload Schedulers – Genesis, Algorithms and Differences

Leszek Sliwko[A], Vladimir Getov[A]

[A] Faculty of Science and Technology, University of Westminster, 115 New Cavendish Street, London W1W 6UW, United Kingdom, {w1518355, V.S.Getov}@westminster.ac.uk

## ABSTRACT

*This paper presents a novel approach to categorization of modern workload schedulers. We provide descriptions of three classes of schedulers: Operating Systems Process Schedulers, Cluster Systems Jobs Schedulers and Big Data Schedulers. We describe their evolution from early adoptions to modern implementations, considering both the use and features of algorithms. In summary, we discuss differences between all presented classes of schedulers and discuss their chronological development. In conclusion we highlight similarities in the focus of scheduling strategies design, applicable to both local and distributed systems.*

## TYPE OF PAPER AND KEYWORDS

Survey: *process schedulers, job schedulers, scheduling algorithms*

## 1 INTRODUCTION

Designing a good scheduler is a complex task. The area of scheduling research is concerned with an effective allocation of available resources with the objective to optimizing one of more performance measures [59]. Depending on the situation, the resources may be CPU time, available memory, I/O operations time-slices, BTS stations in mobile cells network, but also non-IT scenarios, such as managing construction workers working of a building site, doctors located at hospitals, etc. Actually, the first algorithm for the assignment problem was the Hungarian method in 1955 [57], solving the problem of assigning available employees to office jobs based on their skills.

In computer science areas, numerous scheduling algorithms are currently used to determine an effective task/jobs allocation – either on CPU cores or networked nodes. Simple algorithms include: *list scheduling* (LS) assigning jobs from pre-specified list as soon as machine becomes idle [59] *largest processing time first* (LPT) [35], *highest level first* [45] or *round robin* [71] as well as the *weighted round robin* variant [42]. Simple strategies do not require knowledge of unscheduled jobs or all the jobs currently being processed which makes them very popular, especially for online request scheduling [59]. However, not taking into account additional factors such as current server load, network infrastructure or storage availability may result in an inefficient utilization of available machines and overall higher operational system costs.

More complex algorithms rely on the availability of static infrastructure data such as CPU speed, installed memory etc. As an example we find *largest remaining processing time on fastest machine* rule (LRPT-FM), where the job with the most remaining processing time is assigned to fastest machine [44]. LRPT_FM approach offers clear advantages in heterogeneous environments, where the system is composed from machines with different configurations. However this still does not take into account jobs being currently executed. *Fixed-priority pre-emptive scheduling* is an example of a scheduling algorithm commonly used in real-time systems. Early versions of Sun Java Virtual Machine (JVM) implemented this schema, however current versions of JVM use the underlying Operating System thread scheduling model. This scheduling algorithm assumes a hierarchy of task priorities and ensures the processor always executes the highest priority task from those that are currently ready to be executed. This strategy has a serious drawback, as only highest priority tasks are executed – lower-priority tasks could be blocked indefinitely. One solution to this situation is to implement aging, where priority of tasks is gradually increased, ensuring that they will be eventually executed [4].

Besides a local CPU processes allocation, schedulers are commonly used in networked systems. The concept of connecting computing resources has been an active area of research for a considerable period of time. The term 'metacomputing' was established as early as 1987 [73] and since then the topic of scheduling has been one of the key subjects within many research projects – service localizing idle workstations [60] parallel run-time system developed at the University of Virginia [36], blueprints for national supercomputer [37], Globus (1997) metacomputing infrastructure toolkit [28].

As the research showed, the requirements of a load balancer in a distributed system significantly vary from scheduling jobs on a single machine [40]. One important difference are network resources – the machines are usually geographically distributed and transferring data from one machine to another is costly. Additionally, besides effectively spreading jobs across networked machines, the load balancer usually provides a mechanism for fault-tolerance and user session management. Nowadays, load balancers are able to schedule incoming jobs as well as to transfer existing ones to the networked machines.

Scheduling jobs onto parallel nodes is difficult to solve optimally within fixed time, therefore approximation algorithms are usually employed. LS-based techniques have proven to be effective (a variant of LPT algorithm has been shown to have 19/12 guarantee ratio, resulting jobs allocation to optimal jobs allocation in a worst-case scenario) [21][30]. In addition, *Bin Packing* techniques are frequently employed as they naturally share the same decision, i.e. a bin-packing problem is defined as packing a number of items of various sizes into a minimum number of same-size bins) [59]. The heuristic known as *first-fit decreasing* has been shown to have 13/11 ratio guarantee, the worst-case scenario result for optimum items allocation) [83].

In the following chapters we will briefly explain how several current and past schedulers and distributed frameworks work. This will help to develop an understanding of how scheduling algorithms were developed over time and how their concepts have evolved over time. This is by no means a complete survey of all available schedulers, but rather an analysis of some of the landmark features and ideas in the history of schedulers.

In Summary section, we compare all classes of schedulers, with similarities and differences being discussed. In Conclusions, we present a novel approach to categorization of modern workload schedulers. We also suggest a possible unification of the future design of various classes of schedulers.

## 2 OPERATING SYSTEM PROCESS SCHEDULERS

*Operating System Process Scheduler* works within a very short time frames ('time-slices'). During scheduling events an algorithm has to examine planned tasks and assign appropriate CPU times to them [12][68]. This requires schedulers to use simple highly-optimized algorithms with very small overhead. Process schedulers have the difficult task of maintaining a delicate balance between responsiveness (minimum latency) and performance (maximum throughput). This is generally achieved with prioritizing the execution of processes with a higher sleep/processing ratio [65].

Nowadays, the most advanced strategies also take into consideration the latest location (CPU core) where the process actually ran last time ('NUMA (Non-Uniform Access Memory) awareness'), with the aim of reusing the same CPU memory (the level of CPU cache utilization) where possible [7]. This also involves prioritizing in choosing a real idle core first before its logical SMT sibling (aka 'hyperthread awareness'). This is a relatively high data load to examine in a short period of time, thus implementation needs to be strongly optimized for a faster execution.

Operating System Process Schedulers generally provide only a very limited set of tuneable parameters without easy access to modify them. Some of parameters can be changed only during the kernel compilation process (e.g.: compile-time options CONFIG_FAIR_USER_SCHED and CONFIG_FAIR_CGROUP_SCHED) or by a low-level tool *sysctl* (Linux kernel only).

In the following section, we present the most notable process schedulers used in modern operating systems.

### 2.1 Cooperative Scheduling

Early multitasking operating systems (Windows 3.1x, Windows 95, 96 and Me, Mac OS prior to X) implemented a concept known as *cooperative multitasking* or *cooperative scheduling* (CS). In early implementations of CS, applications voluntarily ceded time one to another. Later this was supported natively by the Operating System, although Windows 3.1x used non-pre-emptive scheduler (it did not interrupt the program) and the program needed to explicitly tell the system that it did not need the processor time anymore.

Windows 95 introduced a rudimentary pre-emptive scheduler; however this was for 32-bit applications only [41]. The main issue in CS is the hazard caused by a poorly designed program. CS relies of processes regularly giving up control to other processes in the system. Therefore, if one process consumes all available CPU power, it causes all systems to hang.

### 2.2 Multi-Level Feedback Queue

Perhaps the most widespread scheduler algorithm is *Multi-Level Feedback Queue (MLFQ)*, which is implemented in all modern versions of Windows NT (2000, XP, Vista, 7 and Server), Mac OS X, NetBSD, Solaris and Linux kernels (up to version 2.6, when it was replaced with Q(1) scheduler). It was first described in 1962 in a system known as the Compatible Time-Sharing System [16]. Fernando Corbató was awarded the Turing Award by ACM in 1990 "for his pioneering work organizing the concepts and leading the development of the general-purpose, large-scale, time-sharing and resource-sharing computer systems, CTSS and Multics".

In MHQ jobs are organized into set of queues $Q_0, Q_1, ....$. A job is promoted to higher queue if it does not finish within $2^i$ time units. At any time, algorithm processes the job from the front of the lowest queue. In other words, short processes are given preference [68]. MFQ turns out to be very efficient in practice, while having very poor worst-case scenario [5].

### 2.3 O(n) Scheduler

O(n) Scheduler was used between Linux kernel versions 2.4-2.6, replacing the previously used simple circular queue algorithm [51]. In this algorithm, processor time is divided into epochs. Within each epoch, every task can execute up to its allocated time slice. The time slice is given to each task at the start of each epoch and it is based on the task's static priority added to half of any remaining time-slices from the last epoch [12]. Thus if a task does not use all of its time slice in current epoch, then it can execute longer in the next epoch.

The disadvantage of this approach is relative inefficiency, lack of scalability (especially for multi-core processors) and weakness for real-time systems [51]. The scheduler itself may use a significant amount of time itself if the number of tasks is large (O(n) scheduler requires iteration through all currently planned processes during a scheduling event).

### 2.4 O(1) Scheduler

Between Linux kernel versions 2.6-2.6.23 came the implementation of *O(1) Scheduler*. This design can schedule processes within a constant amount of time (thus the name '*O(1)*'), regardless how many processes are currently running on the kernel [3][81].

The main issue with this algorithm is the complex heuristics used. To mark a task as interactive or non-interactive (interactive tasks are given higher priority in order to boost system responsiveness), O(1) algorithm analyses the average sleep time of process. Those calculations are complex and subject to potential errors, where O(1) may cause non-interactive behaviour from an interactive process [51][65].

### 2.5 Completely Fair Scheduler

At present, Linux kernel implements *Completely Fair Scheduler (CFS)* algorithm (introduced in kernel version 2.6.23) [81]. The main idea behind CFS is to maintain balance ('fairness') in providing processor time to tasks [51], in other words each process should have equal share of CPU time.

CFS implements red-black tree (self-balancing binary search tree structure) holding a queue for future task execution, with spent processor time used as a key and processes with the most sleeping time being prioritized [65]. When the time for tasks is out of balance (meaning that one or more tasks are not given a fair amount of time relative to others), then those out-of-balance tasks should be given time to execute [51].

### 2.6 Brain Fuck Scheduler

*Brain Fuck Scheduler (BFS)* was designed in 2009 and is an alternative to *CFS* and *O(1)* schedulers in the Linux kernel. The main objective of this algorithm was to provide a scheduling strategy suitable for desktop machines (with less CPU cores), that does not require adjustments of heuristic or tuning parameters [39].

In comparison to CFS, algorithm does have lower latency (improves interactivity), but has higher processes turnaround time (lowers performance) [39]. The author does not plan to integrate this scheduler into mainstream Linux kernel (scheduler is available as kernel patch ck1), although there exists several distributions that ship with BFS-enabled kernel, such as Zenwalk, PCLinuxOS, Kanotix and NimbleX.

## 3 CLUSTER SYSTEMS JOBS SCHEDULERS

While responsiveness and low overheads tend to be the focus of process schedulers, it is the case that the role of jobs schedulers is to focus upon scalability and high throughput. Jobs schedulers usually work with queues of jobs spanning to hundreds of thousands and sometimes even millions of jobs [69].

Jobs schedulers usually provide complex administration tools with a wide spectrum of tuneable parameters and flexible workload policies. All configurable parameters can usually be accessed through configuration files or via GUI interface. However, it has been documented that site administrators only rarely stray from a default configuration [23]. The most common scheduling algorithm is simply a First-Come-First-Serve (FCFS) strategy with backfilling optimization.

The common issues cluster schedulers have to deal with are: unpredictable and varying load [63], complex policies, constraints and fairness [23]. Other factors include a rapidly increasing workload and cluster size [47], mixed batch jobs and services [13], legacy software [47], heterogeneous nodes with varying level of resources and availability [76]. There are also issues of hardware malfunctions [23][31] and the detection of underperforming nodes [47].

Another interesting challenge, though rarely tackled in commercial schedulers, is the reduction of total power consumption. Typically, idle machines consume less than half of their peak power [61]. Therefore, the total power consumed by a given Data Centre can be lowered by concentrating tasks on a reduced number of machines and powering down remaining nodes [58][67].

In the following section, we present a few notable industrial-grade job schedulers used in modern distributed computer systems and supercomputers.

### 3.1 Simple Linux Utility for Resource Management

*Simple Linux Utility for Resource Management* (SLURM) is free and Open Source job scheduler for the Linux kernel initially developed for large Linux clusters at the Lawrence Livermore National Laboratory (LLNL). SLURM is used by many of distributed computer systems [82] and supercomputers. TOP500 project, which originated in 1993, ranks and details the 500 most powerful non-distributed computer systems in the world [77] and reports that approximately 50% of world supercomputers are using SLURM as the workload manager.

SLURM uses a *best fit* algorithm based on Hilbert curve scheduling or fat tree network topology and it can scale to thousands of processors [66].

### 3.2 Maui Cluster Scheduler

*Maui Cluster Scheduler* (Maui) is an open source job scheduler for clusters and supercomputers. It has been developed by Cluster Resources, Inc. in early 1990, being currently maintained and supported, but no longer being actively developed by Adaptive Computing, Inc. *Maui* is currently in use at many government, academic, and commercial sites throughout the world on hundreds of IBM SP-2, SGI Origin 2000, and Linux cluster systems [49].

*Maui* implements FCFS strategy [23], with a set of features such as 'advance reservation' (the availability of a set of resources is guaranteed at a particular time), 'backfilling' (optimization allowing shorter jobs to execute while long job at the head of queue is waiting for a free processor [25] and 'fair-share' (when a site administrator can set system utilization targets for users, groups, account, classes and QOS levels [23].

### 3.3 Moab High-Performance Computing Suite

*Moab High-Performance Computing Suite* (Moab) is a direct successor of *Maui* framework. *Moab* has all features from *Maui* and several additional features like basic trigger support, extended policies configuration, graphical administration tools, and a Web-based user portal and better scalability (over 15000 nodes with hundreds of thousands of queued job submissions and over 500 users). Moab currently manages workloads for about 40% of the top 10, top 25 and top 100 on the Top500 list [77] (Young, 2014).

### 3.4 Univa Grid Engine

*Univa Grid Engine* is also known as Oracle Grid Engine, Sun Grid Engine, CODINE (Computing in Distributed Networked Environments), GRD (Global Resource Director) or simply Grid Engine. Univa had acquired it from Oracle in October 2013 [6]. Grid Engine has been developed as an enhancement of CODINE according to requirements from many early customers, such as the Army Research Lab in Aberdeen, and BMW in Munich [32].

Among other features, UGE supports advance reservation, job checkpointing (saving a snapshot of the current application state, which can be used for restarting the application execution in case of a failure [15], Apache Hadoop integration and Amazon EC2 integration for cloud computing. Out of the box, Grid Engine supports two scheduling strategies: FCFS (default) and an optional fair-schare (called 'Equal-Share'), however new strategies can be added, including *the most available* and *lookahead* strategies used to minimize a number of job migrations [79].

In late 2010 after purchase of Sun by Oracle, binaries for version 6.2 update 6 were released without source code. Grid Engine has been forked into multiple open source projects. Currently there are two actively maintained projects: Son of Grid Engine and Open Grid Scheduler.

### 3.5 LoadLeveler

Designed by IBM, *LoadLeveler* manages both serial and parallel jobs over a cluster of servers. *LoadLeveler* implements several scheduling strategies such as plain FCFS, FCFS with backfilling and gang scheduling, simultaneously running a set of related threads or processes on different processors allowing them to exchange messages without sleeping time and context switching [24]. An administrator can rewrite SYSPRIO function to implement alternative strategies [23][53].

LL also supports job checkpointing and it is able to communicate with external schedulers like Maui [23].

### 3.6 Load Sharing Facility

*Load Sharing Facility* (LSF) was created by Platform Computing (acquired by IBM in January 2012 [74] and was based on the Utopia project at the University of Toronto [88].

LSF supports numerous scheduling algorithms like FCFS, fair-share, backfilling and SLA (Service Level Agreements). LSF can also interface with external schedulers like Maui. LSF implements an interesting feature, where a job's priority is gradually increased every time interval (thus the name 'priority escalation'). This scheme results in higher priorities to long-waiting jobs [23].

### 3.7 Portable Batch System

*Portable Batch System* (PBS) was originally developed at NASA Ames research centre under a contract project that began on June 17, 1991. PBS can operate over a huge variety of machines, starting from heterogeneous cluster of loosely coupled workstations to vast parallel supercomputers [8].

PBS includes a number of scheduler strategies, such as FCFS, Shortest Job First (SJF) [68], fair-share and also allows implementation of a custom scheduler in C, TCL or in a specially designed language BaSL [8]. By default SJF strategy is used (starvation is mitigated by marking a particular job as 'starving' and withholding execution of all other jobs until the starving job finishes [23].

There exist three versions of PBS:

- OpenPBS — original open source, suitable for small clusters
- TORQUE — a fork of OpenPBS maintained by Adaptive Computing, Inc.
- PBS Professional — the commercial version of PBS offered by Altair Engineering, Inc.

iPhone/iPad users can also install PBS Express application from Apple Store. PBS Express allows for the monitoring and interaction with a PBS cluster from a smart phone.

### 3.8 Globus toolkit

*Globus Toolkit* is a set of tools for constructing a computing grid. It contains security framework, resource allocation and management strategies, communications libraries, etc. Its origins go back to Supercomputing '95 conference (San Diego, California, USA), where a team of researches build a temporary network of 11 research centres (project 'I-WAY' (Information Wide Area Year)). In order to establish communication between those networks a set of new protocols has been created to allow users to remotely execute applications on computers across the country [28].

Following the success of *I-WAY* experiment, a Defense Advanced Research Projects allocated funds for further research and in 1997 the first version of *Globus Toolkit* was released and soon *Globus Toolkit* was deployed on 80 sites across globe [27].

Occasionally, a computing cluster cannot allocate all resources need by an application at a given time. Application might then wait until the cluster has acquired enough resources, but this will result in bad response time. Alternative strategy is then to co-allocate resources on multiple grid systems and run application that way [72]. In fact, assuming low communication overhead, research demonstrated co-allocation might increase the overall performance of a grid [11][22].

*Globus Toolkit* uses *Dynamically-Updated Request Online Co-allocator* (DUROC) and *Grid Resource Allocation & Management* (GRAM) services to provide all the resources needed by a grid application. Globus Toolkit implements 'gang scheduling'. At first, *DUROC* service decomposes jobs requests into tasks (called 'subjobs') and sends them to remote *GRAM* instance on destination clusters. *GRAM* service communicates with local resource manager (e.g.: Load Sharing Facility, Portable Batch system, Univa Grid Engine, LoadLeveler, Condor, etc.) and allocates resources. When ready (i.e.: resources for tasks have been successfully negotiated), *DUROC* starts a job [72].

*DUROC* service does provide only limited support for job scheduling. It does not implement any jobs queuing mechanism – a jobs submission will simply fail if required resources are not immediately available or their acquisition cannot be successfully negotiated. Also, in situation of single task failure, the whole job is failed and user receives error message [72].

### 3.9 GridWay

*GridWay* is a meta-scheduler developed by the researches at the University of Madrid. It was designed with purpose of providing a flexible and reliable workload manager. *GridWay* was built on top of *Globus Toolkit* framework; therefore it supports a wide range of cluster and grid engines [46].

The code module of *GridWay* system is a *submission agent*. *Submission agent* contains two modules:

- *resource selector* module, which evaluates *requirements* and allocates jobs to hosts (based on *ranking expressions*); both *requirements* and *ranking expressions* are provided by jobs and can be updated dynamically during jobs execution
- *performance evaluator* module, which monitors application's performance in order to detect slowdown and request job's migration to an alternative node

In *GridWay* framework, a job can dynamically modify its *requirements* ('self-adapting') during its execution. An application might initially define a set of minimal *requirements* and keeps updating them later based of its current state. *GridWay* scheduler implements a feature called 'opportunistic migration'. Scheduler periodically evaluates available resources and may detect a better node for a currently executing job (based on dynamically updated *ranking expressions* and *requirements*). Scheduler then evaluates potential benefits of migrating this job to alternative node against the migration overhead [62] and scheduler might migrate the job to better node.

The main drawback of this approach is a need to modify the source code of an application to support this behavior.

### 3.10 HTCondor

*HTCondor* (previously known as *Condor*). The name was changed in October 2012 to resolve a trademark lawsuit [1] and this is the oldest high-throughput software still successfully running today.

*HTCondor* development started at the University of Wisconsin-Madison in 1984 and implemented an idea of stealing idle cycles from university's workstations ('cycle scavenging') [60]. Over years of development, *HTCondor* architecture remained mostly unchanged, while many new features have been implemented and the pool of available nodes grew [76]. Nowadays, *HTCondor* architecture can be used to manage a workload on a cluster system.

A number of tools and frameworks have been built on top of *HTCondor* infrastructure. One example is *DAGMan* (Directed Acyclic Graph Manager). *DAGMan* handles inter-job dependencies, where the programs are nodes (vertices) in the graph and the edges (arcs) identify dependencies. During execution, *DAGMan* orchestrates jobs execution order and schedules jobs directly into the *HTCondor* queue. *HTCondor* then identifies available machines and allocates jobs to them [29].

*HTCondor* implements a 'fair-share' algorithm, where users are allocated machine time based on their priority in the system. Additionally, every user has their own FIFO queue for personal jobs. Condor also supports 'priority pre-emption', where jobs from lower priority users are killed in order to allow higher priority jobs to progress [75].

### 3.11 Mesos

*Mesos* originally began as a research project at University of California, Berkeley [43], but is now hosted in Apache Software Foundation and is being tested at several companies including Twitter and Facebook.

*Mesos* introduces a two-level scheduling mechanism, where a centralized 'Mesos master' acts as a resource manager that dynamically allocates resources to different scheduler frameworks (e.g.: Hadoop, Spark, Kafka, etc.). In case of a master failure *Mesos* uses ZooKeeper framework service to elect a new master [52]. Resources are distributed to the frameworks in the form of 'offers', which contain currently unused resources. Scheduling frameworks have autonomy in deciding which resources to accept and which tasks to run on them [43].

*Mesos* works most effectively when tasks are relatively small (compared to the cluster's size), short-lived and have a high resources 'churn rate' eg - relinquish resources more frequently. In the current design (version 0.20.1 at the time of writing), only one scheduling framework can examine a resource offer at any given time. Therefore, this resource is effectively locked for the duration of a scheduling decision (i.e. concurrency control is pessimistic). A slow decision making scheduler can compromise overall system performance [69].

### 3.12 Open MPI

*Open MPI* is an open source implementation of Message Passing Interface [38] developed and maintained by an international board of high performance computing vendors, academic researchers and applications specialists. *Open MPI* combines a number of libraries, technologies and other resources from a set of projects like LAM/MPI, LA-MPI, FT-MPI and PACX-MPI [31]. It is used in many TOP500 supercomputers such as Roadrunner or K computer [77].

The runtime environment of Open MPI provides a set of services to manage parallel executions in a distributed environment. A set of high-performance drivers is being actively developed for communication channels such as TCP/IP, shared memory, Myrinet, Quadrics and Infiniband. Framework is also transparently capable of handling failures of network devices (when node is equipped with several network devices) [31].

Job scheduling in *Open MPI* is fairly simple and works either on a by-slot basis (selection of all available slots) or by-node basis (selection of all nodes with available slots) round robin schedule. Each *Open MPI* node provides a number of slots available. Frameworks such as SLURM, PBS/Torque and SGE automatically provide an accurate number of slots and if not specified the default value of 1 is used. Each execution of application specifies the number of processes that should be launched (the 'np' switch in mpirun command) and the scheduler then decides where those processes should be allocated. The scheduler takes into account the configured scheduling policy, involving the set of nodes suitable to run processes, defaults and maximum number of slots.

### 3.13 Autopilot

Formerly, the management system for Microsoft's Windows Live Messenger and Live Search services, Autopilot has been expended to support every Windows Live service and well as some other online services such as Windows Live Mail (previously *Hotmail*) resulting in storage space increasing substantially over previous years [48].

The main aim of *Autopilot* is to automate data centre operations and lower the number of people on 24-hour call required to maintain it, therefore lowering capital expense. This is achieved by using more intelligent software to replace much of the repetitive operations handles by data centre staff as well as moving failure management to automated scripts [48].

*Autopilot* provides basic services needed to keep the data centre operational - provisioning and deployment of software, monitoring and hardware lifecycle including repair and replacement. However, job-scheduling policies, such as determining which services should run on which machines are left to individual applications [48].

### 3.14 TORQUE

*TORQUE* (Terascale Open-source Resource and QUEue Manager) is a fork of OpenPBS project maintained by Adaptive Computing, Inc. *TORQUE Resource Manager* provides control over batch jobs and distributed computing resources. In this architecture, the master node runs the pbs_server and the slave nodes run the pbs_mom daemons. Client command interface can be installed on any host (not necessary on system node).

In default configuration, the simple FIFO job scheduler is deployed on the master node. The job scheduler interacts with pbs_server daemon allocate nodes to jobs based on configured resource usage policies. *TORQUE* users can choose to use an alternative scheduler such as *Maui* or *Moab*.

At the time of writing, Czech National Grid Infrastructure *MetaCentrum* is evaluating an experimental extension to *TORQUE Resource Manager* [56], where an ad-hoc jobs placement mechanism has been replaced by a more sophisticated planning-based approach. This strategy will allow users to see when and where their jobs will be executed and predict behavior of the cluster system [14].

In the new approach, the constructed schedule is periodically (every 5 minutes) evaluated and incrementally improved by a Tabu Search algorithm (algorithm's runtime is bounded by 2 seconds in each iteration) [55]. Various metrics such as makespan, slowdown, response time or wait time may be used as optimizations criteria [14].

Research also points to an interesting fact of notorious inaccuracy of job's runtime estimations, due to a need to prevent job being killed due to an overrun. Jobs in fact often complete earlier than expected and this phenomena results in cumulative gaps and unnecessary high waiting times [54]. As demonstrated such 'corrupted schedule' can be immediately fixed by re-running optimization routine and 'compressing schedule' [14].

Initial experiments show very promising results in comparison to various backfilling strategies [14].

### 3.15 Borg and Omega

To support its operations, Google utilizes a high number of data centres around the world (at the time of writing, Google has 12 data centres [2]. To orchestrate all its jobs in such a complex environment, Google has been using a custom job-scheduling system unofficially known as *Borg* [13].

Google's *Borg* is effectively a *monolithic* scheduler. It uses a single, centralized scheduling algorithm for all jobs. In contrast, t*wo-level or dynamic* schedulers such as *Mesos* or *TORQUE* have a single resource manager that makes a resource offers to multiple independent scheduler instances. However, regardless of various optimizations acquired over years including internal parallelism and multi-threading, to address head-of-line blocking and scalability problems, Google decided to rewrite a scheduler as part of project *Omega* [13].

The concept behind Omega is to deploy several schedulers working in parallel. The scheduler instances are using a share state of available resources, however the resource offers are not locked during scheduling decisions (optimistic concurrency control). In case of conflict, when two or more schedulers allocated jobs to the same resources, all involved jobs are returned to the jobs queue and scheduling is re-tried [69].

This approach seems to be rather successful as shown in a study. It eliminates head-of-line job blocking problems and offers better scalability, however it also generates additional overhead for solving resource collisions. Nevertheless, the better scalability benefits often outweigh the incurred additional computations costs and scalability targets have been achieved [69].

A point to note is that similar to *MetaCentrum* users, Google cluster users tend to overestimate memory resources needed to complete their jobs to prevent jobs being killed due to an exceeding of allocated memory. In over 90% of cases, users tend to overestimate the amount of resources that they require, wasting in some cases close to 98% of the requested resource [63].

# 4 BIG DATA SCHEDULERS

*Big Data* is a term given to the storage and processing of any collection of data sets so large and complex that it becomes unrealistic to process using traditional data processing applications (generally based on relational database management systems). *Big Data* also applies to statistics and visualization packages.

Due to size *Big Data* is generally difficult to work with. Analyzing large data sets requires parallel software running on huge farms of servers [50] and that introduces new challenges of managing the workload and optimizing the usage of a cluster. It depends on the individual organization how much data will be called *Big Data*, but the following examples may be considered to get an idea of scale:

- New York Stock Exchange produces about one terabyte of new trade data per day
- Facebook hosts approximately 10 billion photos and currently about one petabyte of storage
- The Large Hadron Collider (Geneva, Switzerland) produces about 15 petabytes of data per year [80]

*Big Data* systems tend to be more specialized in their design, usually tackling only a very limited set of problems [47]. They often provide their own api [80][86] and sometimes even the custom programming language, as seen with Skywriting in *CIEL* [64].

Despite these limitations *Big Data* systems are relevant to this research, as jobs scheduling and performance optimization remain common challenges.

*Big Data* frameworks have a dual purpose, storing system data on its nodes (usually three replicas of each data block are used for fault-tolerance purposes [33][80] and secondly, to process this data via parallel tasks using the same nodes. A common optimization is applied, namely 'data locality', where a scheduler attempts to schedule tasks near the data blocks required.

Recently, many specialized frameworks have been created. Below, we will discuss and assess some of the most interesting and important, providing a brief description and focusing on job scheduling aspect in each.

## 4.1 Dryad

Dryad is a general-purpose framework for execution of data-parallel applications in distributed computers network developed at Microsoft Research. The project had several preview releases, but was ultimately dropped in October 2011 [26] and Microsoft shifted focus to the development of *Hadoop*.

The development of an application for *Dryad* is modeled as a *directed acyclic graph* (DAG) model. The developer defines an application dataflow model and supplies subroutines to be executed at specified graph vertices. The developer has also a fine-control over the communication protocols (e.g.: files, TCP pipes, shared memory) used in graph. The result is a developing style similar to 'piping' in Unix bash utilities (i.e.: streaming output from one tool to another, such as: cat file.txt | grep 'word' | wc –l), but in distributed flavor [47].

In comparison to *MapReduce*, *Dryad* features a much more lower-level programming schema, but all parallelization and task allocation and distribution mechanisms are effectively hidden from the developer. Therefore user does not need to have understanding of concurrency control mechanisms such as threads, semaphores or locks.

The *Dryad's* scheduler keeps a record of state and history of each vertex in a graph. A vertex might be re-executed in case of a failure (e.g.: node hardware malfunction) and more than one instance of a given vertex might be executed at the same time, meaning execution is versioned to avoid conflicts among runs. Upon a successful execution, one version is selected and returned as a result [47].

A vertex or any pre-defined channels might have preferences and constraints for the node it is to be run [47] and this allows the developer to implement a very basic 'location optimality', where task and input data are forced to be located on the same machine.

The *Dryad's* job scheduler implements Greedy strategy. In this approach the scheduler assumes that currently scheduled job is the only job running on a cluster and always selects the best node available. Tasks are run by remote daemon services, which periodically update the job manager about vertex's execution status. If any task has failed more than a configured number of times, the entire job is marked as failed [47].

### 4.2 MapReduce

At the time of writing, *MapReduce* is the most widespread adopted principal for processing large sets of data in parallel. The name *MapReduce* originally referred only to the Google technology, developed in 2003 [19] and patented in [20] to simplify building of the inverted index for handling searches at Google.com, however the term is now widely used to describe a wide range of software (e.g. Hadoop, CouchDB, Infinispan, MongoDB, Riak, etc.) based on this concept.

The concept behind *MapReduce* was first presented in 2004 [18] and was inspired by 'map' and 'reduce' operations (hence the name 'map-reduce' or *MapReduce*) present in Lisp and many other functional programming languages [34].

However, the key contributions of *MapReduce* are the scalability and fine-grained fault-tolerance. This means that a failure in the middle of a multi-hour execution does not require restarting the job from scratch achieved for a variety of purposes by optimizing the execution engine once [19]. 'Map' is an operation that performs filtering and sorting all key-value pairs from the input data set, while 'reduce' performs summary operations (e.g.: counting the number of rows matching specified condition/s, yielding fields frequencies). 'Map' operation is used in the first step of computation by being applied to all available data. Due to its nature, 'map' can be executed in parallel on multiple machines (i.e. on distributed data set) and it is highly scalable. In the next step, the job goes into an intermediate state in which the framework gathers all values returned by 'map' workers. Then, a 'reduce' operation is fed with all received values supplied using an iterator, thus allowing the framework to process list which may not fit into machine memory [34].

### 4.3 Hadoop

Following publication of *MapReduce* concept [18], Yahoo! engineers have started Open Source project Hadoop. February 2008, Yahoo! announced that it's production search index was being generated by a 10000-core *Hadoop* cluster [80]. Subsequently, many other major Internet companies (e.g.: Facebook, LinkedIn, Amazon, Last.fm [84]) have joined the project and deployed it in their architectures [34]. *Hadoop* is currently hosted in Apache Software Foundation as Open Source project.

*Hadoop* runs on top of a *Hadoop Distributed File System* (HDFS, similar to Google's implementation of *MapReduce*, which runs on Google File System [33]. Users submit *MapReduce* jobs, which consist of 'map' and 'reduce' operations/ implementations. *Hadoop* splits each job into multiple 'map' and 'reduce' tasks, which then process each block of input data (typically 64MB or 128MB).

Since its first release, *Hadoop* acquired a number of optimizations:

*A.* 'locality optimization' was introduced, where the scheduler allocates 'map' task to the closest possible node to the input data required by it (the following allocation order is used: the same node, the same rack or finally the remote rack [85].

*B.* *Hadoop* also uses 'backup tasks', where a speculative copy of a task is run on a separate machine to finish computation faster. If the first node is available, but behaving poorly (such node is called 'straggler', this behaviour can arise from many reasons such as faulty hardware or misconfiguration), a job would be as slow as the misbehaving task [84]. Google estimated that using 'backup tasks' can improve job response times by 44% [18].

Currently, *MapReduce* in Hadoop comes with a selection of schedulers:

#### 4.3.1 FIFO scheduler

Early versions of *Hadoop* had a very simple default scheduling system where the user jobs were scheduled using a simple FIFO queue with five priority levels [85].

Typically, jobs were using whole cluster, so jobs had to wait their turn. When then job scheduler was choosing the next job to run, it was selecting jobs with the highest priority, thus it could result in low-priority jobs being delayed endlessly. Alternatively, as FIFO scheduler was not supporting pre-emption, a high-priority job could be blocked by a long-running low-priority job that started before the high-priority job was added to schedule [80].

#### 4.3.2 Fair Scheduler

*Fair Scheduler* (together with *Capacity Scheduler* described in the next section) is part of a cluster management technology *YARN* (Yet Another Resource Negotiator) framework, which is one of the key features in the second-generation Hadoop 2 version [78].

*Fair Scheduler* focuses on giving each cluster user a fair share of cluster resources over time, thus creating an illusion for each user of owning a private cluster [85]. Each user has their own pool of jobs and scheduler uses a version of 'max-min fairness' [9] with minimum capacities guarantees (specified as the number of 'map' and 'reduce' task slots) to allocate tasks across pools. As more jobs are submitted, free tasks slots are given to the jobs in such a way as to give each user a fair share of the cluster computation capacity [80]. Thus, in busy environment, submitting more jobs by a user will not result in more cluster resources being used by this user. When one pool is idle (not using minimum share of his task slots), other pools are allowed to use available task slots.

Jobs pools can have variable weights configurations. Usual scenario is to create one pool per user and special pools for production jobs [85].

*Fair Scheduler* also supports pre-emption, thus if any given job pool is running over its capacity, its tasks will be killed to make free slots for under-running job pools [80].

### 4.3.3 Capacity Scheduler

In *Capacity Scheduler*, a cluster is made up of a number of FIFO queues. Those queues might be hierarchical (a queue might have children queues) and each queue has allocated task slots capacity (separate for 'map' and 'reduce' tasks). Task slots allocation between queues is similar to sharing mechanism between pools as seen in *Fair Scheduler* [80].

Essentially Capacity Scheduler can be seen as a number of separate *MapReduce* clusters with FIFO scheduling for each user or organization.

### 4.4 HaLoop

*HaLoop* framework has been developed on top of *Hadoop* in response to the poor performance of the former when running iterative jobs (Bu et al., 2010). The reason for this behaviour is a default mechanism, where *Hadoop* writes the output of each MapReduce job to the distributed file system and reads it back on during the next iteration [34].

By adding various caching mechanisms and optimizations and making framework loop-aware (e.g. adding programming support for iterative application, storing the output data on the local disk), iterative jobs performance has been massively improved. *HaLoop* reduces query runtimes by 1.85, and shuffles only 4% of the data between mappers and reducers [10].

HaLoop's scheduler keeps a record of every data block processed by each task on physical machines and tries to schedule subsequent tasks taking inter-iteration locality into account [10]. This feature helps to minimize costly remote data retrieval, thus tasks can use data cached on a local machine.

### 4.5 Spark

Similarly to *HaLoop*, Spark's authors noted a suboptimal performance of iterative *MapReduce* jobs in *Hadoop* framework [86].

*Spark* is built on top of HDSF, however it does not follow the two-stage model of *MapReduce*. Instead it introduces *resilient distributed datasets* (RDD) and parallel operations on these datasets:

- 'reduce' - combines dataset elements using a provided function
- 'collect' - sends all the elements of the dataset to the user program
- 'foreach' - applies a provided function onto every element of a dataset

Additionally, this framework provides two types of shared variables:

- 'accumulators' - variables onto each worker can apply associative operations (therefore they are efficiently supported in parallel)
- 'broadcast variables' - sent once to every node, with nodes then keeping a read-only copy of those variables

For certain kind of applications (e.g. iterative machine learning algorithms and interactive data analysis tools), *Spark* outperforms *Hadoop* in order of magnitude, while retaining the scalability and fault tolerance of *MapReduce* [86].

Spark job scheduler implementation is conceptual similar to *Dryad's*, however it takes into account which partitions of RDD are available in memory (framework re-computes missing partitions) and tasks are sent to the closest possible node to the input data required ('locality optimization') [87].

Another interesting feature implemented in *Spark* is a concept of 'delayed scheduling'. When a head-of-line job that should be scheduled next (according to fairness) cannot launch a local task, it lets other jobs start their tasks instead and repeatedly re-tries. However, if the job has been skipped long enough (typically up to 10 seconds), it launches non-local task. As typical *Spark* workload consists of short tasks (i.e. it has a high task slots churn), tasks have higher chance to be executed locally (there is no cost of retrieving the input data from a remote node).

This feature helps to achieve almost optimal 'data locality' with a minimal impact on fairness and the cluster throughput might be increased by up to two times (analysis has been performed on Facebook's workload traces) [87].

### 4.6 CIEL

Designed and implemented at the Cambridge Computer Laboratory (University of Cambridge), *CIEL* is a universal execution engine for distributed computation. *CIEL* implements master-slave architecture, where the master is responsible for coordinating and dispatching tasks to workers. Workers execute tasks and also can store result state to objects, which might be an input to following tasks. This data can be directly exchanged between workers by making remote calls to workers' store objects [64]. A task can dynamically spawn 'children' tasks, effectively delegating the production of its output to its children [70].

*CIEL* uses its own scripting language for writing CIEL jobs – 'Skywriting'. Skywriting is a full Turing-complete language, which allows developers to mix task creation and data-dependent control flow [64].

*CIEL* is an open-source project licensed under the BSD license. Author also implemented *CIEL* variant running on multi-core machines (i.e. HTTP transport mechanism replaced by shared memory and communication between tasks re-implemented with OS-level pipes and semaphores) [70].

| Scheduler class | Resources requirements known pre-execution | Fault-tolerance mechanisms | Configuration | Common algorithms | Scheduling decision overhead | Design focus (aside performance) |
|---|---|---|---|---|---|---|
| Operating System Process Schedulers | No | No | Simple (compile-time and runtime kernel parameters) | MLFQ, O(n), O(1), CFS, BFS (with locality optimization) | very low - low | • single machine<br>• responsiveness<br>• simple configuration |
| Cluster System Job Schedulers | Yes | Yes | Complex (configuration files and GUI) | FCFS (with backfilling and gang-scheduling), SJF | low - high | • distributed nodes<br>• fairness<br>• complex sharing policy<br>• power consumption |
| Big Data Schedulers | No[(1)] | Yes | Complex (configuration files and GUI) | FIFO (with locality optimization and gang-scheduling), Fair Scheduler, Greedy | low - medium | • specialized frameworks<br>• parallelism<br>• distributed data storage |

(1) *MapReduce* jobs tend have consistent resource requirements (i.e.: in majority of cases, every *map* task processes roughly the same amount of data (input data block size is constant), while reduce task requirements shall be directly correlated to the length of returned data)

Table 1: Schedulers comparison

### 5 SUMMARY

In this survey we presented a number of available schedulers from early implementations to modern versions. It may be noted that each class of scheduler started with a simple job queue and developed over time as specific sets of problems emerged:

*A. Operating System Process Schedulers* evolved with focus on maximizing responsiveness [65], while still providing good performance. While CPU switches between processes in a very rapid manner, the modern operating system process scheduling algorithms were designed with a very low overhead [81]. The majority of end-users for operating system process schedulers are non-technical; therefore those schedulers usually have a minimum set of configuration parameters [39]. Introduction and popularization of multi-core processors by Intel and AMD in early 2000s (i.e.: Intel

Core Duo and AMD Phenom II X2), enabled applications to execute *in parallel* and Operating Systems Schedulers started developing in similar direction as distributed systems schedulers. Modern Operating System Process Schedulers also implement cache 'locality optimization' when deciding which CPU core the task will be allocated to.

*B. Cluster Systems Jobs Schedulers* have a difficult mission of ensuring 'fairness' [9] (i.e.: sharing cluster resources proportionally to every user) while maintaining stable throughput in a very dynamic environment. Cluster systems usually allow administrators to implement complex resource sharing policies with multiple input parameters. Cluster systems usually implement fault-tolerance strategies (i.e.: 'checkpointing' [15]) and sometimes also focus on minimizing power consumption [58]. Surprisingly, the most popular approach to scheduling is simple *First-Come-First-Served* strategy with variants of backfilling. However, due to rapidly increasing cluster size, current research focuses on parallelization, as seen with models such as Google's Omega [15]. Cluster users are notorious in overestimating resources needed for completion of their tasks, which results in cluster system job schedulers often over-allocating resources [54][63].

*C. Big Data* systems are still rapidly developing. Nodes in Big Data systems fulfil the dual purpose of storing distributed file system parts (e.g.: Google File System [33], its successor Colossus [17] or Hadoop Distributed File System [34]) and providing a parallel execution environment for system tasks. Job schedulers in this class inherit general design from cluster system's jobs schedulers, but are usually very specialized for the purpose of a framework and focused on 'locality optimization' or running a given task on a node where input data is stored or in the closest proximity to it.

## 6 CONCLUSIONS

The design of modern scheduling strategies and algorithms is a challenging and evolving field of study. While early implementations were often based on very simplistic approaches such as a circular queue (also known as 'Round Robin'), it is the case that modern solutions use complex load balancing schemas (i.e.: Google's Omega, where multiple schedulers are working in parallel and competing for resources [69]) and introduce concepts like previously mentioned 'fairness', 'checkpointing', 'backup tasks', etc.

During this research, we have noted many similarities between scheduling strategies used in all classes of schedulers. Early Operating System level schedulers were focused primarily on responsiveness and performance [16][68]. However, their design focus changed dramatically with introduction of multi-core processors and modern scheduler implementation strategies supporting parallel execution [81] and similar to those designed for distributed Cluster systems.

The experiments with CIEL scheduler demonstrate that strategies used in allocating tasks in a distributed system of nodes can be adapted to effectively work on the Operating System level [70]. A recent introduction of *Completely Fair Scheduler* (based on a model of 'fairness') to Linux kernel (since kernel version 2.6.23) also highlights current trends of mixing concepts from both local and distributed systems when designing scheduling strategies.

In the future, we expect research on all classes of schedulers to be increasingly more joined and design combining ideas from both the Operating System level as well as distributed Cluster systems.